\documentstyle[psfig,astrobib,times]{mn-ab}

\def\spose#1{\hbox to 0pt{#1\hss}}
\def\la{\mathrel{\spose{\lower 3pt\hbox{$\sim$}}
        \raise 2.0pt\hbox{$<$}}}
\def\ga{\mathrel{\spose{\lower 3pt\hbox{$\sim$}}
        \raise 2.0pt\hbox{$>$}}}

\def\H0{{\it H}$_0$}
\def\Ms{{\it M}$_\odot$}

\def\q0{{\it q}$_0$}

\def\kmpspMpc{km~s$^{-1}$~Mpc$^{-1}$}
\def\Ms{{\it M}$_\odot$}

\def\ergpspsqcm{erg~s$^{-1}$~cm$^{-2}$}

\def\Zs{$Z_{\odot}$}

\def\d{{\rm d}}
\def\ie{i.e.\ }
\def\eg{e.g.\ }

\def\bgunit{keV~cm$^{-2}$~s$^{-1}$~sr$^{-1}$~keV$^{-1}$}

\title[The soft X-ray background]
{The soft X-ray background: evidence for widespread disruption
  of the gas halos of galaxy groups}
\author[K. K. S. Wu et al.] 
{\parbox[]{6.5in} {K. K. S. Wu,$^{1,4}$ A. C. Fabian$^1$ and
P. E. J. Nulsen$^{2,3}$}\\
\\
$^1$ Institute of Astronomy, Madingley Road, Cambridge CB3 0HA\\ 
$^2$ Department of Engineering Physics, University of Wollongong, Wollongong
     NSW 2522, Australia\\
$^3$ Harvard-Smithsonian Center for Astrophysics, 60 Garden Street,
Cambridge MA 02138, USA\\
$^4$ kwu@ast.cam.ac.uk}
\date{}

\begin{document}
\maketitle
\begin{abstract} 
  Almost all of the extragalactic X-ray background (XRB) at 0.25 keV
  can be accounted for by radio-quiet quasars, allowing us to derive
  an upper limit of 4 \bgunit\ for the remaining background at 0.25
  keV.  However, the XRB from the gas halos of groups of galaxies,
  with gas removal due to cooling accounted for, exceeds this upper
  limit by an order of magnitude if non-gravitational heating
  is not included. We calculate this using simulations of halo
  merger trees and realistic gas density profiles, which we require to
  reproduce the observed gas fractions and abundances of X-ray
  clusters. In addition, we find that the entire mass range of groups,
  from $\sim 5\times 10^{12}$ to $\sim 10^{14}$\Ms, contributes to the
  0.25 keV background in this case.
  
  In a further study, we reduce the luminosities of groups by
  maximally heating their gas halos while maintaining the same gas
  fractions.  This only reduces the XRB by a factor of 2 or less.  We
  thus argue that most of the gas associated with groups must be
  outside their virial radii. This conclusion is supported by X-ray
  studies of individual groups. 
  
  The properties of both groups and X-ray clusters can be naturally
  explained by a model in which the gas is given excess specific
  energies of $\sim 1$ keV/particle by non-gravitational heating.
  With this excess energy, the gas is gravitationally unbound from
  groups, but recollapses with the formation of a cluster of
  temperature $\ga 1$ keV. This is similar to a model proposed by Pen,
  but is contrary to the evolution of baryons described by Cen \&
  Ostriker.

  The excess energy is most likely injected by galaxies in the
  smaller `branches' of the halo merger tree ($\la 5\times 10^{12}$\Ms), by
  active galactic nuclei and possibly supernovae. The heating process
  may therefore play an important part in the evolution of galaxies.

  In addition to the soft XRB spectrum, we simulate source counts
  in two bands: 0.1--0.4 keV and 0.5--2 keV, for comparison with
  present and future data.

\end{abstract}

\begin{keywords}
  galaxies: clusters: general -- galaxies: haloes -- intergalactic
  medium -- cooling flows -- X-rays: galaxies
\end{keywords}

\section{Introduction}

Compared to X-ray clusters, relatively little is known about the hot
gas halos of galaxy groups (henceforth simply `groups').  With the
Chandra and XMM satellite missions we can expect much to be revealed
about them.  However, the extragalactic soft X-ray background (XRB)
below $\sim 1$ keV already provides a useful probe of their mean
properties.  For example, we shall show that (in the absence of
non-gravitational heating) the 0.25 keV background probes almost the
entire mass range of groups ($\sim 5\times 10^{12}$--$10^{14}$\Ms,
corresponding to temperatures of $T\sim 0.1$--1 keV).  

Groups are cosmologically important, for a majority of the baryons in
the universe may be associated with them at the present day
\cite{fhp98}. In addition, groups connect galaxies to clusters in a
hierarchical merger tree. This suggests that a large fraction of the
heavy elements and `excess energy' (due to non-gravitational heating)
injected into the inter-galactic medium (IGM) would have to pass
through groups before ending up in X-ray clusters. However, it has
been shown that strong heating of the IGM is required to satisfy
constraints on the soft XRB \cite{pen99}.  The heating required by
Pen, $\delta T\sim 1$ keV, would imply that most of the gas would not be
gravitationally bound to the potential wells of groups, contrary to
the scenario described above.  It is also interesting that this amount
of heating is similar to that required to explain the properties of
X-ray clusters \cite{wfn98,loewe99,pen99,wfn99}. Together they
therefore suggest a consistent picture of heating in the IGM.
Furthermore, the mechanism for heating the IGM is likely to
have significant implications for the evolution of galaxies (Wu et
al., 1999; hereafter WFN99), whether the heat source be supernovae or
active galactic nuclei (AGN).

In this paper, we examine in more detail the contraints on groups,
using simulations of halo merger trees and drawing on elements of our
semi-analytic model of galaxy formation (WFN99). One important
difference from Pen's model is our inclusion of cooling, in particular
its role in removing gas from halos. Our methods for calculating the
XRB also differ significantly. Broadly speaking, the main questions we
aim to address are whether groups are significantly affected by
non-gravitational heating, and if so, what range of groups are
affected.

Although a large fraction of the soft XRB below about 0.8 keV
originates in our own Galaxy, the extragalactic component can be
measured with shadowing experiments.  As reviewed by \citeN{wr98},
measurements of the extragalactic XRB in the 0.1--0.4 keV band
(hereafter 0.25 keV) appear to be converging on a value in the range
20--35 \bgunit. However, the bulk of the extragalactic XRB is due to
AGN. A significant fraction of the XRB at 1 keV has been resolved into
AGN. By modelling the QSO X-ray luminosity function and its evolution,
\citeN{bsgsg94} estimated the contribution of QSOs to the 1--2 keV XRB
(see also \citeNP{schmi98}).  Their results correspond to 35--55 per
cent of the 1--2 keV background measured by \citeN{gendr95}.  Now,
\citeN{lfewm97} find that radio-quiet quasars have a mean spectral
slope of $\alpha=-1.72\pm 0.09$ over the range 0.2--2 keV, though they
obtain a significantly flatter slope for (less common) radio-loud
quasars. 
Therefore, if we
assume that radio-quiet quasars account for 90 per cent of the quasar
contribution to the 1 keV background,
then they contribute at least $0.9\times 0.35$ or about 30 per cent of
the 1 keV background. Using the 1 keV background measured by
\citeN{gendr95}, which is 9.6 \bgunit, we can thus obtain a lower
limit to the contribution of radio-quiet quasars below 1 keV.  At 0.25
keV the lower limit is 31 \bgunit.  From the range quoted by Warwick
\& Roberts, it follows that gas halos almost certainly contribute less
than 4 \bgunit at 0.25 keV, as most of the 0.25 keV background is
already accounted for by QSOs. We shall adopt this value as an upper
limit to the contribution from gas halos.  (We note that Boyle et al.,
Laor et al.\ and the detections quoted by Warwick \& Roberts all
obtained their data from the same instrument, namely the ROSAT
Position Sensitive Proportional Counter.)  The main assumption we have
made lies in the application of the mean properties measured by Laor
et al.\ to all quasars. On the other hand, we have used the lowest
expected contribution of QSOs at 1 keV, along with the highest
expected value for the 0.25 keV background.
We do not attempt here to assign a confidence level to this limit
since the main uncertainties are systematic and not statistical.

The models in this paper are intended to be realistic, but simple
enough for comparisons to be easily made. We consider two low-density
universes with $\Omega_m=0.3$: one with no cosmological constant and
the other with $\Omega_\Lambda=0.7$. A Hubble constant of $H_0=100h$
\kmpspMpc\ where $h=0.5$ is assumed (the insensitivity of our results to
$h$ is discussed in section~\ref{seccalib}).
We use CDM-type fluctuation power spectra, normalized to
match the observed abundance of X-ray clusters.  In this way, the
abundance of groups are in effect extrapolated from the X-ray cluster
abundance.  Likewise, in our fiducial models we assume that groups
have a gas fraction of $0.06h^{-3/2}$ when they virialize, as this is
the mean value measured for clusters \cite{evrar97,ef98}. This is a
reasonable choice since groups are the progenitors of clusters. Gas
halos are assumed to be isothermal and hydrostatically supported
in Navarro, Frenk \& White (1997; NFW) potential wells. The
resulting gas profiles are found to closely model X-ray clusters
\cite{mss97,ef98}. In the absence of excess energy, the model gas
profiles of groups and clusters are self-similar to a good
approximation. This is a reasonable assumption to make (see
hydrodynamic simulations such as Navarro, Frenk \& White
1995)\nocite{nfw95}, especially as we are looking for large
differences between model results and observations.  We synthesize
spectra using the MEKAL spectral synthesis code \cite{kaast92}.

This paper is organised as follows: in section~\ref{secsim} we
describe our model, followed by how we calculate the XRB and source
counts. We simulate the XRB from 0.05--2 keV, and calculate source
counts in two bands: 0.1--0.4 keV and 0.5--2 keV.  In most of
sections~\ref{secsim} and \ref{secfid} we assume that
non-gravitational heating is absent.  However, we also describe
simulations where halos are required to follow observed
luminosity-temperature relations extrapolated to lower temperatures.
The fiducial results are discussed in section~\ref{secfid}. In
section~\ref{secfur} we investigate the effects of including
non-gravitational heating, and of allowing the gas fraction of halos
to be determined by inheritance.  We discuss some implications of our
results in section~\ref{secimp} and summarize our conclusions in
section~\ref{secconc}.

\section{Simulation} \label{secsim}

The main components in the calculation of our fiducial results are the
halo merger tree, the gas density profiles of virialized halos, and
the spectral synthesis model.

The halo merger trees are simulated using the \citeN{ck88} block
model, as in our earlier models (\eg WFN99). The
smallest regions simulated have a mass of $1.5\times 10^{10}$\Ms. We
assume a collapse hierarchy of 20 levels, so that the mass of the
largest block is $2^{19}\times 1.5\times 10^{10} = 7.9\times
10^{15}$\Ms. This is the total mass of the region simulated in one
`realisation' of the merger tree.

The mean density of collapsed halos are specified to conform exactly
with the spherical collapse model.  In open cosmologies without a
cosmological constant, the mean density of a virialized halo is
assumed to be $18\pi^2\approx 178$ times the background density of an
Einstein-de-Sitter universe of the same age.  In flat cosmologies with
a cosmological constant, we use the analytic approximation given by
Kitayama \& Suto (1996; equation A6)\nocite{ks96}, where the mean
density of a halo is given by
\begin{equation}
  \rho_{\rm vir}=18\pi^2(1+0.4093(\Omega_m^{-1}-1)^{0.9052}) \overline{\rho}_b.
\end{equation}
Here, the density parameter, $\Omega_m$, and the background density,
$\overline{\rho}_b$, are evaluated at the time of collapse.  At very
early times, results from the two prescriptions tend to the same
value, as the vacuum density is then small relative to the density of
collapsing halos. However, at late times the simple
prescription for open cosmologies becomes a poor approximation in
$\Lambda$-cosmologies. For example, for a halo that virializes today
in a cosmology given by $(\Omega_m,\Omega_\Lambda,h)=(0.3,0.7,0.5)$, the
first prescription underestimates the mean density by about 20 per
cent.  Since the X-ray luminosity of a gas halo scales roughly as the
density squared, this difference can be significant.
From the mass of the collapsed halo, the radius of the halo,
$r_{\rm vir}$, thus follows.

The second component in the calculation is the assumed gas density
profile of halos. To simplify the synthesis of spectra, we consider only
isothermal gas halos. We assume the gas to be in hydrostatic
equilibrium in \citeANP{nfw97} (1997; NFW) potential wells. In other
words, the total density of the halo is described by the NFW profile:
\begin{equation}
  \rho(r)={\delta_c \rho_{\rm crit}\over (r/r_s)(1+r/r_s)^2},
\end{equation}
where $\rho_{\rm crit}=3H^2/8\pi G$ and $H$ is the Hubble parameter at
the time of collapse.  The parameter $\delta_c$ is calculated as
described in the Appendix of NFW. The scale radius $r_s$ is then
uniquely determined by the mean density calculated above.
The concentration parameter,
$c$, is defined to be $r_s/r_{\rm vir}$. (This differs slightly from
NFW, as we use a more detailed derivation for the mean density. In
particular, their equation~(2) that relates $\delta_c$ and $c$ is
slightly modified in this model.) From the NFW profile, the
gravitational potential is given by
\begin{equation}
     \phi(x) = \alpha\left( -\frac{\ln(1+x)}{x} + \frac{1}{1+c} \right),
\end{equation}
where $x=r/r_s$ and $\alpha = 4\pi G \rho_s r_s^2$. It then follows
that a gas halo of temperature $T$ in hydrostatic equilibrium takes
the form
\begin{equation}
  \rho_{\rm g}\propto (1+x)^{\eta/x},
\end{equation}
where $\rho_{\rm g}$ is the gas density, and $\eta=\mu m_{\rm
  H}\alpha/(kT)$ (Wu et al., 1998; WFN98). Here, the
mean mass per particle of the gas is denoted by $\mu m_{\rm H}$, and
$k$ is the Boltzmann constant. This gas profile closely approximates
the conventional $\beta$-model if $\beta=\eta/15$ \cite{mss97}, and
models most X-ray clusters very well. For large clusters the mean
value of $\eta$ is observed to be about 10.5 \cite{ef98}.

We calculate $\eta$ by first specifying the total energy of the gas
halo, which then uniquely determines $\eta$.  This is described in
more detail elsewhere (WFN98; Model B in WFN99). To summarize, in the
absence of excess energy, the total specific energy of the gas halo 
(thermal plus gravitational) is
required to be proportional to the specific gravitational energy of
the entire halo; we then calibrate this energy relation by matching to
the largest clusters. Although halos described by the NFW profile are
not exactly self-similar (as the concentration $c$ varies), this
relation for the gas halos expresses one form of self-similarity.  For
the purposes of this paper, the main point to note is that this
results in $\eta$ close to 10.5 for all gas halos. In addition, the
resulting gas temperatures are well-approximated by $T_{\rm SIS}$, the
temperature that the gas would have if both gas and dark matter had
power-law density profiles: $\rho\propto r^{-2}$; \ie $kT_{\rm
  SIS}/(\mu m_{\rm H})=GM_{\rm tot}/(2r_{\rm vir})$, where $M_{\rm
  tot}$ is the total mass of the halo. In general, $T$ scatters
between $T_{\rm SIS}$ and $1.05T_{\rm SIS}$ for clusters, with the
upper end of the range increasing to $1.1T_{\rm SIS}$ as we go down to
halo of $\sim 10^{12}$\Ms.

We assume that isolated galaxies have halo masses of up to $\sim
10^{12}$\Ms. Since we are primarily concerned with the properties of
halos, in this paper the term `group' simply refers to a halo of
mass greater than a few $10^{12}$\Ms\ (but less than $\sim
10^{14}$\Ms).

In our fiducial models, we assume that all halos have a gas mass
fraction of 0.17 when they virialize. This is the mean cluster gas
fraction measured by \citeN{evrar97} and \citeN{ef98} (using $h=0.5$),
within a radius $r_{500}$ defined to be such that the mean density
within $r_{500}$ is $500\rho_{\rm crit}$. We note that the gas
fraction of both observed and simulated clusters tend to increase with
the radius within which they are measured. Since $r_{500}<r_{\rm
  vir}$, it is therefore likely that the true gas fractions of
clusters within $r_{\rm vir}$ are higher than this. This strengthens
our argument below as it would increase our fiducial estimates of the
XRB.

\subsection{Calculating the X-ray background}

We now describe in detail how we calculate a mean spectrum for the
X-ray background.  Spectra are simulated using the MEKAL model
\cite{kaast92}, to which we supply two parameters: the gas
temperature, $T$, and the metallicity, $Z$, in units of the solar
abundance, \Zs\ \cite{ag89}.  From the spectrum given by MEKAL, we
calculate the cooling function, $\Lambda(T,Z)$. In this way, the model
self-consistently estimates the cooling time of the gas.

\subsubsection{Cold and hot collapses} \label{sectau}
We recall that in the collapse of less massive halos, the
cooling time of the gas, $t_{\rm cool}$, can be shorter than the
free-fall time to the centre of the halo, $t_{\rm ff}$. In this case
the gas cools fast enough that it is not hydrostatically supported,
and in spite of any shock heating, the gas temperature remains well
below the virial temperature most of the time. We refer to this
as a cold collapse (WFN99). In our model, gas is labelled as cold when
$\tau=t_{\rm cool}/t_{\rm ff}<\tau_0$, where $\tau_0\sim 1$ is a free
parameter. Otherwise, the gas is labelled as hot. This is important
when calculating the X-ray background, as we expect cold gas to
radiate little, if at all, in the X-ray band. Therefore
we only integrate contributions from hot gas and use a fiducial
value of $\tau_0=1$.

For isothermal gas profiles, $\tau$ is almost always a monotonically
increasing function of radius (see Appendix A of WFN99 for a more
detailed discussion).  This allows us to define a radius, $r_{\rm
  CF}$, such that gas outside $r_{\rm CF}$ is labelled as hot and gas
inside as cold. As halo mass increases, $r_{\rm CF}$ moves from
outside the virial radius to the centre of the halo.  This transition
is fairly abrupt and occurs over about one decade in mass, in halos of
$\sim 10^{12}$\Ms.  
For halos in the transition region, $r_{\rm CF}$ is found by solving
the equation $\tau(r_{\rm CF})=\tau_0$ numerically.

(In our model, the transition to the hot-gas regime is made more
abrupt by supernova feedback from star formation.  We assume that cold
gas rapidly forms stars, which lead quickly to type II supernovae. If
a sufficient fraction of a collapse is cold, then the energy from
supernova feedback is able to eject the rest of the atmosphere,
including the hot gas \cite{nf97}. By using a much lower value of
$\tau_0=0.1$, we show that these complications at the transition
region have a very small effect on the predicted XRB and source
counts, and do not affect our conclusions in any way.)

When a hydrostatically-supported hot gas halo occurs, any gas that
cools is assumed to form low-mass stars or `baryonic dark matter', in
analogy to cooling flows in X-ray clusters. A possible mechanism for
low-mass star formation in cooling flows is described by \citeN{mb99},
for the case of elliptical galaxies. It remains possible (if not
likely) that some normal star formation and feedback occurs in cooling flows.
However this does not affect our main conclusions (the effects of strong
heating are investigated in section~\ref{secfur}).

\subsubsection{Spectral synthesis} \label{secsyn}
We calculate model XRB spectra from $E=0.05$ keV to 2
keV. We first divide this range into equal logarithmic bins of width
$\Delta\log_{10} E=0.01$.  Suppose photons in the universe belonging
to the energy bin $(E,E+\Delta E)$ at the present day have a number
density of $\Delta n$, then the corresponding energy flux per
steradian (\ie the intensity) is given by $E \Delta n c /(4\pi)$,
where $c$ is the speed of light (in practice, we use the geometric
mean of $E$ and $E+\Delta E$ in place of $E$ in this expression).
Dividing by $\Delta E$ thus gives the intensity per unit energy, which
we express in units of keV~cm$^{-2}$~s$^{-1}$~sr$^{-1}$~keV$^{-1}$.
Below, we refer to these bins as `collecting bins'.

Since each realisation of a merger tree simulates a region of constant
comoving volume, the photon density $\Delta n$ is simply given by the
total number of photons (of the correct energy) emitted in the
simulation divided by the present-day volume of the simulation. 

We now consider a gas halo at some redshift $z$.
Given $T$ and $Z$, its {\em rest-frame} spectrum is
calculated with MEKAL, also in equal logarithmic bins of width
$\Delta\log_{10} E=0.01$.  The spectrum is integrated to give the
cooling function $\Lambda(T,Z)$---defined such that the
bolometric luminosity per unit volume is given by $n_e n_{\rm H}
\Lambda(T,Z)$, where $n_e$ and $n_{\rm H}$ are the electron and
hydrogen number densities respectively. We then calculate the number
of photons emitted in each `rest-frame bin' per unit total energy that
is radiated.  These ratios are used throughout the remaining
calculation.

For gas radiating at a redshift of $z$, each {\em collecting} bin is
blueshifted accordingly (from $E$ to $(1+z)E$) to collect photons of
the correct energy.  Photons from each rest-frame bin are then dumped
into the nearest collecting bin.
Since the amount of blueshift is a continuous
function of $z$, the correspondence between rest-frame bins and
collecting bins shifts by one bin each time $(1+z)$ decreases by 0.01
dex. This occurs $\sim 10$ times over the life of a halo (which is the
period from one collapse to the next in the block model---we
interprete collapses as major mergers).

To summarize, for each halo we calculate the total energy
radiated during each time interval between bin shifts. From this the
number of photons deposited in each collecting bin is found. The total
number of photons collected in the simulation then gives the XRB
spectrum as explained above.  

\begin{figure}
\centerline{\psfig{figure=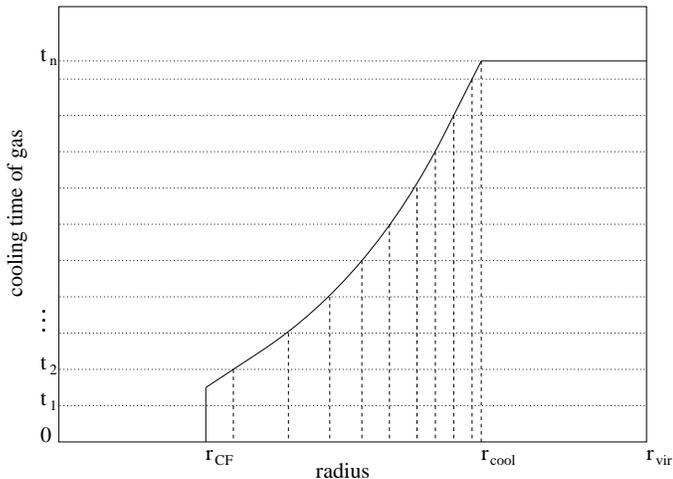,width=0.5\textwidth}}
\caption[Schematic diagram of how the total energy radiated by a halo is
  computed]
  {Schematic diagram of how the total energy radiated by a halo is
  computed. The solid curve encloses the gas emitting in X-rays at any
  given time: between $r_{\rm CF}$ and $r_{\rm cool}$ it gives the
  cooling time of the gas, but gas outside $r_{\rm cool}$ does not
  have time to cool before the next collapse, which occurs at $t_n$.
  Gas inside $r_{\rm CF}$ is labelled as cold and therefore does not
  contribute to the XRB. The correspondence between rest-frame bins
  and collecting bins shifts by one position at the times
  $t_1,\ldots,t_{n-1}$. Thus the energy radiated during each time
  interval needs to be calculated. For the gas between $r_{\rm CF}$
  and $r_{\rm cool}$, our method involves first dividing the halo into
  shells, as shown by the dashed lines (see text).}
\label{figdemo}
\end{figure}

In what follows we describe how we calculate the energy radiated by a
halo, taking into account the effects of cooling.  The most general
case is illustrated in Fig.~\ref{figdemo}, which shows the amount of
gas contributing to the XRB as a function of time.  Also shown are the
times $t_i$ ($i=1,2,\ldots,n-1$) when bin shifts occur, measured from
when the halo virializes. $t_n$ gives the time of the next collapse.

Gas inside $r_{\rm CF}$ is regarded as cold and therefore is not
included in the calculation.  Between $r_{\rm CF}$ and the cooling
radius, $r_{\rm cool}$, the cooling time is shorter than the time to
the next collapse: gas in this region is assumed to cool out once its
cooling time has elapsed. The cooling time is estimated by
\begin{equation} \label{tcooleqn}
  t_{\rm cool}={3\over 2}{\rho_{\rm g}
                   kT/\mu m_{\rm H}\over n_e n_{\rm H} \Lambda(T,Z)}.
\end{equation}
As explained in WFN99, the gas profile
that we use to estimate such quantities must be regarded as notional,
as it cannot describe the gas halo at all times. In particular, the
gas halo redistributes itself as gas in the centre cools out.
Nevertheless, the gas profile allows us to estimate the behaviour of
different subsets of gas in the halo. 

Between $r_{\rm cool}$ and $r_{\rm vir}$, gas does not have time to
cool before the next collapse. Here, the
energy radiated during each time interval is easily calculated from
its bolometric luminosity, which is given by $\int_{r_{\rm
    cool}}^{r_{\rm vir}} n_e n_{\rm H} \Lambda(T,Z) \d V$.  The number
of photons dumped into each collecting bin thus follows.

This is less straightforward to apply to the gas inside $r_{\rm
  cool}$, because the amount of gas changes continuously with time.
However, the gas inside $r_{\rm cool}$ cools completely, so that the
energy radiated is simply given by $(3/2)NkT$, where $N=\int \rho_{\rm
  g}/(\mu m_{\rm H}) \d V$ is the number of particles that cool.  To
estimate the amount of energy radiated in each time interval, we first
divide the halo into shells separated by the radii $R_{\rm
  cool}(t_i)$. The function $R_{\rm cool}(t)$ gives
the radius where cooling time is equal to $t$. For the shell
with an outer radius of $R_{\rm cool}(t_i)$, we assume that
$(t_j-t_{j-1})/t_i$ of the energy it radiates (as given above) is emitted in
the $j$-th time interval. In this way, the total energy radiated is
correctly accounted for, but the `allocation' of energies between time
intervals is less precise. However, the latter is equivalent to
determining how the photons are binned, and the binning of photons is
already uncertain by $\pm 1$ bin, because the bin shifts are uncontinuous.
It follows that the derived spectrum is `smeared'
by $\sim 1$ bin width. 

Note that the release of $(3/2)kT$ per particle that cools must be
regarded as a lower limit: in reality, the weight of the overlying gas
is likely to at least maintain the pressure of the gas as it cools, so
that gravitational work raises the total energy radiated to $(5/2)kT$
per particle or more in most cases. In addition, the density of the
gas increases as it cools and moves to smaller radii, so that the
luminosity of a given parcel of gas increases. As a result, $(3/2)kT$
leads to a reasonable estimate of the cooling time. For the gas
outside $r_{\rm cool}$, the density also increases as the gas moves
inwards.  Therefore, our estimate of the energy released in this case
should also be a lower limit.

A caveat of our model is that any intrinsic absorption that might
occur in the cooling region (in analogy to 
cluster cooling flows) is ignored. Since this is confined to
the cooling region at any given time, while most of the simulated XRB
arises from outside that region, its effect is unlikely to be important.

\subsection{Source counts}


The $\log N$--$\log S$ function for X-ray halos is calculated in two
bands: the 0.1--0.4 keV band, for comparison with future results from
Chandra and XMM, and the 0.5--2 keV band, for comparison with results
from the Wide-Angle ROSAT Pointed X-ray Survey (WARPS; Scharf et al.\ 
1997)\nocite{schar97} and the slightly deeper counts made by
\citeN{rdbng95}.

Since the simulated merger trees provide no spatial information, it is
assumed that the halos are distributed randomly in space. The
probability of a given source in the simulation being observable in
principle (with an infinitely sensitive telescope) depends on how long
it `exists'.  Suppose a source exists for a short period
corresponding to $\d z$.  The comoving volume observable by us on the
entire sky in this redshift interval is given by
\begin{equation} \label{comovevol}
  \d V_c = {c\over H_0} {[D_L(z)/(1+z)]^2 \over
 \sqrt{\Omega_m(1+z)^3 + (1-\Omega_m-\Omega_\Lambda)(1+z)^2 +
 \Omega_\Lambda} }
                   4\pi|\d z|,
\end{equation}
where $D_L(z)$ is the luminosity distance (given below). Suppose we
create an infinite universe by tiling together copies of the same
simulation. Then the mean number of copies of this particular source
that we expect to see is given by $\d V_c$ divided by the volume of
simulation. This is therefore the contribution of this source to the
mean source count on the whole sky.

We integrate source counts for fluxes of $S=10^{-16}$--$10^{-13}$
erg~s$^{-1}$~cm$^{-2}$, which we divide into logarithmic bins of width
$\Delta \log_{10} S=0.1$. The flux from a gas halo is given by $S=L_{\rm
  X}/(4\pi D_L^2)$, where $L_{\rm X}$ is the luminosity in the
relevant, blueshifted band.  For $\Omega_\Lambda=0$, $D_L$ is given by (\eg
Peacock 1999)\nocite{peaco99bkc}:
\begin{equation}
  D_L = (1+z) {2c\over H_0} 
          {\Omega_m z + (\Omega_m-2)(\sqrt{1+\Omega_m z}-1) \over
            \Omega_m^2 (1+z)},
\end{equation}
but for $\Omega_m+\Omega_\Lambda=1$, $D_L$ is integrated numerically:
\begin{equation}
  D_L = (1+z) {c\over H_0} \int_0^z {\d z' \over 
     \sqrt{\Omega_m(1+z)^3 + \Omega_\Lambda}}.
\end{equation}

Equation~\ref{comovevol} is integrated over the lifetime of a given
halo in the simulation, to give the mean number of copies of that halo
that we expect to see on the sky. To do this we associate $\Delta z$
with the same time intervals described in the previous section.  This
gives a good approximation for the cosmologies and redshifts that we
are interested in. Luminosities and fluxes are calculated at the
beginning of each time interval. Each interval then contributes to the
number count in the relevant flux bin as described above.  (Using the
average luminosity in each time interval only changes the number counts
by $\sim 5$ per cent.)

We tested our code by summing all the discrete sources to give an
alternative calculation of the XRB at a given energy. To do this we
replaced the above energy bands with one very narrow band at this
energy, and integrated the total flux in this band.  The resulting
intensity per keV agreed well with the spectrum derived using our
usual method.

\subsection{Empirical calculations based on $L_{\rm X}-T$ relations}
\label{secemp}

It is well-known that self-similar gas halos (such as those described at the
beginning of this section) do not match the observed properties of
X-ray clusters.  Substantial heating is required to lower the
luminosities of the smaller clusters, in order to match the observed
relation of $L_{\rm X}\propto T^3$ (WFN98,99; Pen 99). How this
extends to groups is less clear. A clue is offered by the $L_{\rm
  X}-T$ distribution of the Hickson Compact Groups or HCGs \cite{pbeb96},
which is significantly steeper than that for clusters. However, this
only extends down to about 0.5 keV, and the scatter in the
distribution appears to be intrinsically very large.  (In
section~\ref{secfur}, we propose a reason for the large scatter.) The
steeper $L_{\rm X}-T$ relation for groups suggests that the effects of
heating are even more severe than in clusters.  Other X-ray studies of
groups \cite{mdmb96,hp99} confirm this view.

Since we only require the temperature and bolometric luminosity of a
gas halo in order to estimate its spectrum, we have performed separate
simulations based on (extrapolations of) observed $L_{\rm X}-T$
relations. A basic difficulty with the $L_{\rm X}-T$ relation for
groups is that a large fraction of groups do not have detectable X-ray
emission, so that an X-ray selected sample is likely to give a biased
estimate of the average luminosity. In fact, optically-selected
samples also face difficulties, as they may include chance projections
of galaxies.  On top of this, the X-ray temperature of a group can
differ significantly from the temperature associated with the velocity
dispersion $\sigma$ of the galaxies \cite{pbeb96,hp99}.  If we suppose
that $\sigma$ reflects the underlying dark matter distribution, then
the gas temperature in the self-similar case is approximately given by
$kT/(\mu m_{\rm H})=\sigma^2$.  In reality, the $T-\sigma$
distribution for groups exhibits a large scatter about the best-fit
power law, which has a different slope from the self-similar relation.
For the HCGs, the best-fit power law yields temperatures (as a
function of $\sigma$) than are up to a factor of 2 higher than the
self-similar prediction \cite{pbeb96}.

Faced with these uncertainties, we have simply applied the observed
$L_{\rm X}-T$ relations and assumed that $T$ is the same as the
temperature given in the self-similar model.  In addition, we assume
that the luminosity of a halo remains constant during its lifetime.
This is reasonable because the time-averaged luminosity of a halo in
the simulation should correspond to the observed luminosities averaged
over many groups. In this way, we are able to calculate XRB spectra
and source counts as described above.  Although crude, the results
give an insight into what the true composition of the soft XRB might
look like.

The best-fit $L_{\rm X}-T$ relation for the HCGs is given by
\cite{pbeb96}:
\begin{equation}
  \log L_{\rm bol}= (43.17\pm 0.26) + (8.2\pm 2.7) \log T,
\end{equation}
where $L_{\rm bol}$ is the bolometric luminosity in erg~s$^{-1}$ and
$T$ is in keV. As noted above, we also apply this relation below its
observed range. Despite the uncertainty in the slope, it is clearly
much steeper than the relation for clusters.  For clusters, we use the
best-fit given by \citeN{wjf97}:
\begin{equation}
  \log L_{\rm bol}= (42.7\pm 0.1) + (2.98\pm 0.11) \log T.
\end{equation}
This spans a range of $T\approx 2$--10 keV.  The two relations
intersect at $T=0.8$ keV. Thus we are able to apply the cluster relation above
0.8 keV, and the group relation for lower temperatures. We refer to
this combination as `PW'. We also perform simulations with only the
cluster relation, extrapolated to all temperatures. We refer to these
as `W' simulations.

In the simulations, we do not include scatter in the above relations.
This is permissible for estimating the XRB, which depends on the mean
properties of halos. However scatter can significantly affect the
resulting source counts. For our purposes, this is more easily
adjusted for after the simulation, and we discuss it with the results.

\section{Fiducial results} \label{secfid}

\subsection{Calibration} \label{seccalib}

Perhaps the two most important `parameters' in this study are the gas
fraction and the mass function of the relevant halos.

For the fiducial models discussed in this section, we set the initial
gas fraction of all halos to be 0.17.  Although this may be
unrealistic in detail, it provides a simple point of reference.  As
for the mass function of groups, the primary constraint, albeit
indirect, is the mass function of X-ray clusters at $z=0$.  We require
all of our simulations to reproduce the temperature function of X-ray
clusters, for which we use the best-fit power law measured by
\citeN{esfa90}.  (The temperature function is a reasonably direct way
of constraining the mass function of clusters, whereas the luminosity
function is sensitive to other properties, such as the gas fraction
and the structure of gas halos.)  It follows that the abundance of
less massive halos are in effect extrapolated from the observed
cluster abundance, by assuming CDM power spectra. To investigate the
inherent uncertainty in this, we consider two different values for the
CDM shape parameter $\Gamma$.

As in WFN99, we assume a baryon density parameter of $\Omega_b=0.02
h^{-2}=0.08$ \cite{bt98,bntt99}. Following \citeN{sugiy95}, the CDM
shape parameter is then given by $\Gamma=\Omega_m h
\exp(-\Omega_b(1+\sqrt{2h}/\Omega_m))=0.106$. 
Using this value of $\Gamma$ in an OCDM universe with
$(\Omega_m,\Omega_\Lambda,h)=(0.3,0,0.5)$, we are able to match the
temperature function of X-ray clusters using $\sigma_8=0.75$ (as shown
in WFN99).

For a $\Lambda$CDM cosmology with
$(\Omega_m,\Omega_\Lambda,h)=(0.3,0.7,0.5)$ and the same values for
$\Omega_b$ and $\Gamma$, we are able to reproduce the cluster
temperature function using $\sigma_8=0.8$. Both of the above results
for $\sigma_8$ agree remarkably well with Eke et al.\ (1998,
and private communication for $\Lambda$CDM)\nocite{ecfh98}.

For each cosmology, we also consider a higher value of $\Gamma=0.25$,
as measured by \citeN{pd94} from galaxy surveys. In both cosmologies,
we find that the same values of $\sigma_8$ are able to reproduce the
cluster temperature function, in agreement with Eke et al.\ (1998).

\begin{table}
  \caption{List of simulations and legend for
    Figs.~\ref{figopenspec}--\ref{figlambdaspec} and 
    Figs.~\ref{figopencns}--\ref{figlambdasns}. Each line in the table 
    corresponds to two simulations, performed in the OCDM and 
    $\Lambda$CDM cosmologies. Only the parameters
    that change between simulations have been listed. The `W' and `PW'
    simulations are described in section~\ref{secemp}.}\label{tabsims}
  \begin{tabular}{lllll}
 line-style & metallicity (\Zs) & $\tau_0$& $\Gamma$ & $L_{\rm X}-T$ fixed?\\
    \hline
    solid             & 0.3    & 1     & 0.106    & no   \\
    dotted            & 0.03   & 1     & 0.106    & no   \\
    dashed            & 0.3    & 1     & 0.25     & no   \\
    dot dash          & 0.3    & 0.1   & 0.106    & no   \\
    dash dot dot dot  & 0.3    & 1     & 0.106    & yes, W  \\
    long dashes       & 0.3    & 1     & 0.106    & yes, PW  \\
    \hline
  \end{tabular}
\end{table}

A list of simulations is given in Table~\ref{tabsims}, which we repeat
for each cosmology. The table gives the values of parameters that vary
between simulations. We assume the same metallicity for all gas
halos in any given simulation.  For each simulation that assumes no
non-gravitational heating (`no' in the last column), we use 100
realisations of the merger tree.  These simulations produce similar
results to each other.  Thus, to test that their results have
converged, we increased the number of realisations in one of these
simulations by a factor of 4.  This made no noticable difference to
the results.

In the simulations based on extrapolations of observed $L_{\rm X}-T$
relations (`yes' in the last column), the XRB is dominated by more
massive halos, so that 400 realisations are required to obtain
convergence.

Finally, we note that our results are not sensitive to increases in
$h$.  For our fiducial simulations that assume self-similarity, this
can be understood by noting that the model is normalized to reproduce
the contribution of the largest observed clusters (which are close to
self-similar, \citeNP{af98x}) to the XRB. Since that contribution is a
directly measured quantity, it has no dependence on $h$. The rest of
the simulated XRB is then a non-trivial extrapolation of the cluster
contribution. A similar argument applies to all our simulations. 

To test the sensitivity, we also simulated the first model in
Table~\ref{tabsims} using $h=0.7$, without changing $\Gamma$ and using
$\Lambda$CDM. The gas fraction of clusters scales as $h^{-3/2}$, but
this is compensated by the number density of halos---as expressed by
the temperature function---which scales approximately as $h^3$.
The resulting XRB spectrum is about 10 per cent higher and the source
counts (in the given flux range) are virtually unchanged.

\subsection{The XRB spectrum from hot gas halos}

\begin{figure}
\centerline{\psfig{figure=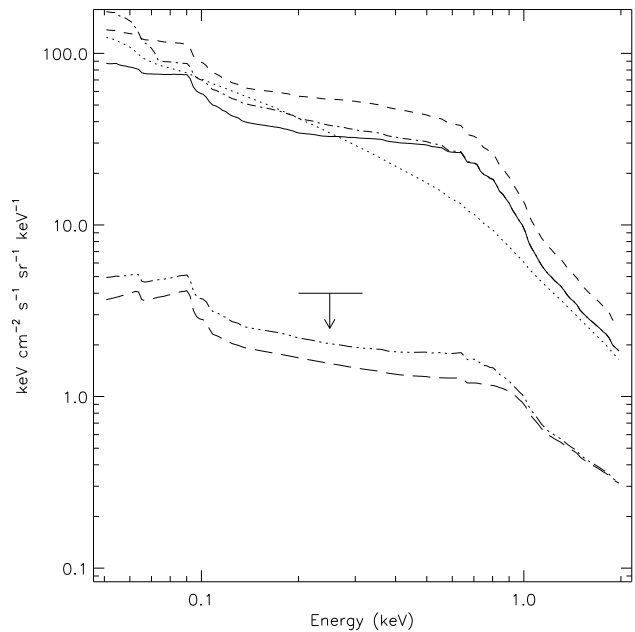,width=0.5\textwidth}}
\caption{Simulated XRB spectra in the OCDM cosmology. The legend
  for the different lines is given in Table~\ref{tabsims}.  The upper
  limit for the gas halo contribution to the observed 0.25 keV
  background is shown by the arrow.}
\label{figopenspec}
\end{figure}

\begin{figure}
\centerline{\psfig{figure=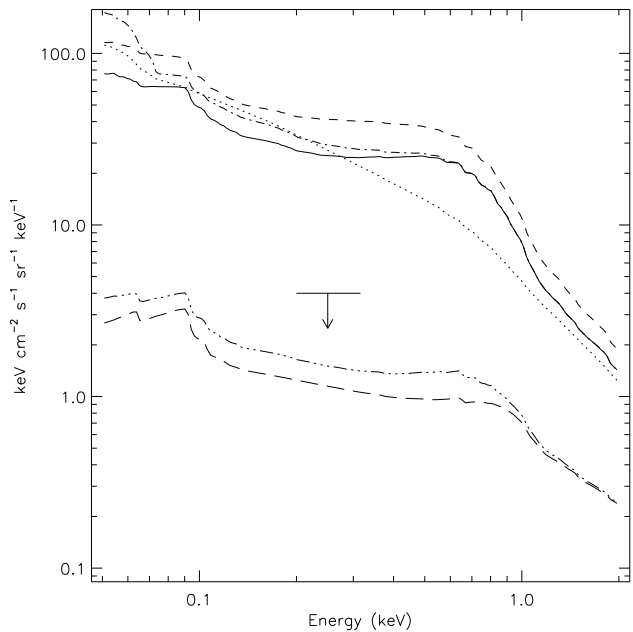,width=0.5\textwidth}}
\caption{As Fig.~\ref{figopenspec}, but in the $\Lambda$CDM cosmology.}
\label{figlambdaspec}
\end{figure}

The XRB spectra from the simulations listed in Table~\ref{tabsims} are
displayed in Figs.~\ref{figopenspec} and \ref{figlambdaspec}, for OCDM
and $\Lambda$CDM respectively.  It is clear from the two figures that
the two cosmologies produce similar results. (The OCDM spectra are
only slightly higher than the $\Lambda$CDM spectra, by up to a factor
of 1/3 depending on the energy.) The 0.25 keV backgrounds predicted by
the simulations that assume no heating are all much higher than our
upper limit of 4 \bgunit\ (section~1)---several by over an order of
magnitude. These simulations also overpredict the 1 keV background,
which should be about 1 \bgunit; the extragalactic background at 1
keV is about 10 \bgunit\ \cite{gendr95}, but only about 10 per cent of
this is likely to be from groups and clusters \cite{mchar98}.

To illustrate the importance of cooling, we modified one of these
simulations as follows to switch off cooling.  
For all halos that contribute to the XRB in the fiducial simulation,
a) we label all of their gas as hot (thus $r_{\rm CF}=0$),
and b) we suppose that no gas is removed as a result of cooling. This
increases the 0.25 keV background by a factor of 5, and increases the
1 keV background by almost 3-fold. These large increases show that if
cooling is not included, then gas will often end up radiating many
times its thermal energy. (In this case, our results
increase to a level comparable to the 0.25 keV background calculated
by Pen (1999), using clumping factors obtained from hydrodynamic
simulations.)

We shall now discuss the differences between the above spectra.  We
regarded the parameters used to obtain the solid spectra as the
`default' parameters (Table~\ref{tabsims}) and varied each of the parameters in
turn to observe the consequences.  Reducing the metallicity from 0.3
to 0.03 \Zs\ (dotted spectra) reduced the amount of line-emission, so
that the resulting XRB spectrum is much smoother. By comparison, a large
`bump' in the solid spectrum at around 0.7 keV can be seen.
It corresponds to the (redshifted) iron L complex. However, the
changes at the other energies are more modest; the spectrum is almost
independent of metallicity at 0.25 keV. It is interesting that the XRB
spectrum {\em increases} at certain energies when we reduce the
metallicity. This occurs in parts of the spectrum where line-emission
is weak. It is possible because the cooling time of gas increases when
we reduce the metallicity.
Increasing the CDM shape parameter to $\Gamma=0.25$ (dashed spectra)
raises the power at sub-cluster scales, resulting in more galaxy
groups. In both cosmologies, this raises the XRB spectrum by roughly
50 per cent.

\begin{figure}
  \centerline{\psfig{figure=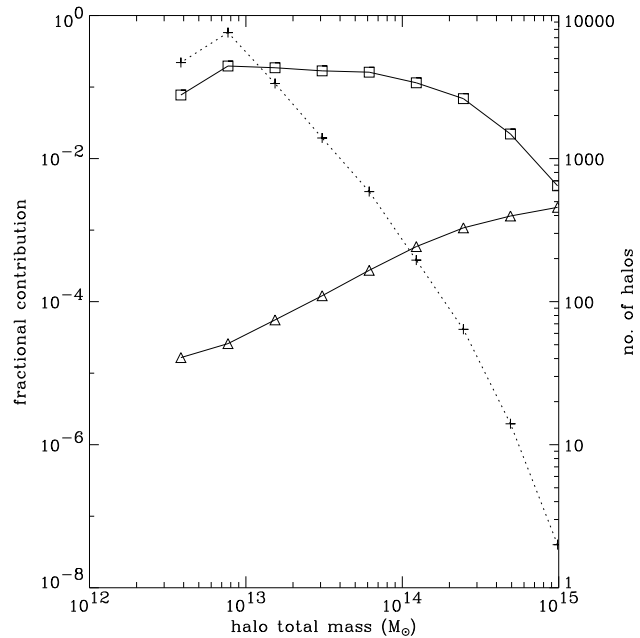,width=0.5\textwidth}}
\caption{Plot of halo contributions to the XRB at 0.25 keV, for the
  case of the solid spectrum in the $\Lambda$CDM cosmology
  (Fig.~\ref{figlambdaspec}). The solid
  lines correspond to the left axis, and the dotted line to the right
  axis. The squares give the fractional contribution from all halos of a
  given mass, and the triangles give the mean contribution from those
  halos. The squares in effect give the contribution per unit
  logarithmic interval in mass. The crosses show the number of hot gas
  halos obtained in the simulation; thus the squares are the product
  of the triangles and crosses.}
\label{figlambda2cpho}
\end{figure}

\begin{figure}
\centerline{\psfig{figure=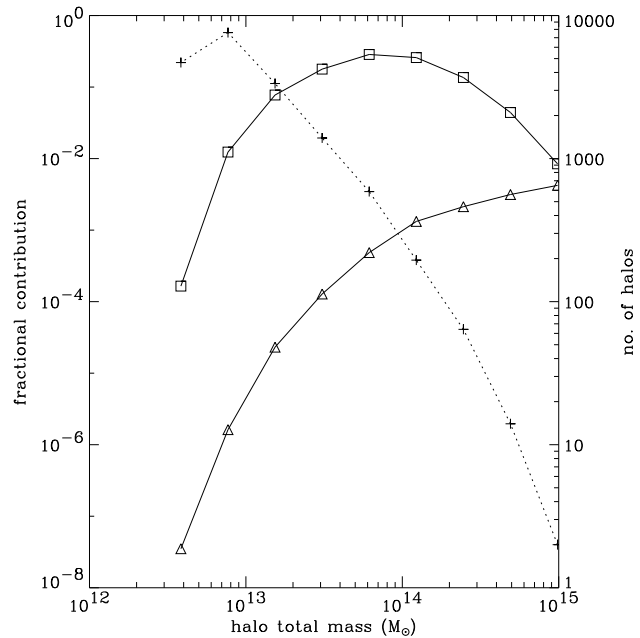,width=0.5\textwidth}}
\caption{As Fig.~\ref{figlambda2cpho}, but for the XRB at 1 keV.}
\label{figlambda2spho}
\end{figure}

To better understand the implications of these results, we need to
know which types of halo are contributing to the soft XRB.  We
therefore computed the fractional contribution of halos belonging to
each mass in the block model. This was done at two energies: 0.25 keV
and 1 keV, which correspond to the two energy bands used in our source
counts. In Figs.~\ref{figlambda2cpho} and \ref{figlambda2spho} we show
the fractional contributions to the solid spectrum in the $\Lambda$CDM
case (Fig.~\ref{figlambdaspec}), at 0.25 keV and 1 keV respectively.
The corresponding plots for the OCDM case are almost the same, and the
results for the dotted and the dashed spectra are very similar.  In
these plots, the solid lines correspond to the axis on the left and
the dotted line to the axis on the right. The main result is given
by the squares, which show the fraction of the XRB due to all halos of
a given mass (since the masses increase by factors of 2, this is
equivalent to the contribution per unit logarithmic interval in mass).

At 0.25 keV (Fig.~\ref{figlambda2cpho}), the squares form a relatively
flat plateau from $\sim 5\times 10^{12}$\Ms\ to $\sim
10^{14}$\Ms. In other words, the 0.25 keV background is contributed by
almost the entire range of halos corresponding to groups.  We assume
that isolated galaxies have total masses of up to $\sim 10^{12}$\Ms,
and that an X-ray cluster with a temperature of 1 keV has a mass of
$\sim 10^{14}$\Ms.  This result means that if we wish to match our
upper limit on the 0.25 keV background, then we must reduce the
contributions of almost all groups. 

The triangles in Fig.~\ref{figlambda2cpho} give the mean fractional
contribution per halo as a function of mass. The plus signs
give the total number of hot gas halos obtained in the simulation.
Therefore, the product of these two curves reproduce the flat plateau
traced by the squares.

The situation is quite different for the XRB at 1 keV.  In
Fig.~\ref{figlambda2spho} the squares peak quite sharply at around
$6\times 10^{13}$\Ms. The difference is brought about by the much
steeper curve traced by the triangles, because less massive halos
contribute less at 1 keV. 

\begin{figure}
\centerline{\psfig{figure=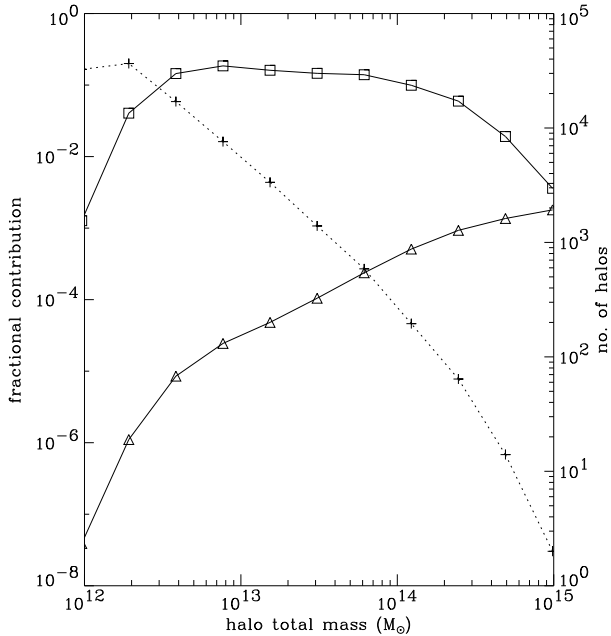,width=0.5\textwidth}}
\caption{As Fig.~\ref{figlambda2cpho}, but for the dot-dashed spectrum at
  0.25 keV in the $\Lambda$CDM cosmology. Lowering $\tau_0$ to 0.1
  results in hot gas halos at lower masses. In addition, the low-mass
  drop-off in the squares is now due to the sharp decline in
  contribution per halo (shown by the triangles), which should be
  contrasted with Fig.~\ref{figlambda2cpho}.}
\label{figlambda4mcpho}
\end{figure}

Returning to Fig.~\ref{figlambda2cpho}, we note that the sharp drop in
contributions below $4\times 10^{12}$\Ms\ appears to be entirely due
to the low-mass `cutoff' in the dotted curve (cf.\ 
Fig.~\ref{figlambda2spho}). This `cutoff' results from the transition
from cold to hot collapses, which is controlled by the parameter
$\tau_0$. Therefore, to investigate the sensitivity of the XRB to
$\tau_0$, we reduced its value from 1 to 0.1. This gave the dot-dashed
spectra in Figs.~\ref{figopenspec} and \ref{figlambdaspec}, which show
only a small increases from the solid spectra (about 25 per cent at
0.25 keV).  For the $\Lambda$CDM case, we show the new breakdown of
contributions at 0.25 keV in Fig.~\ref{figlambda4mcpho}.  The curves
now extend to a lower mass, showing that the plateau does indeed end
around $4\times 10^{12}$\Ms, even though the transition from cold to
hot collapses has been pushed to less massive halos. This explains why
the 0.25 keV background is not sensitive to the parameter $\tau_0$.

\begin{figure}
\centerline{\psfig{figure=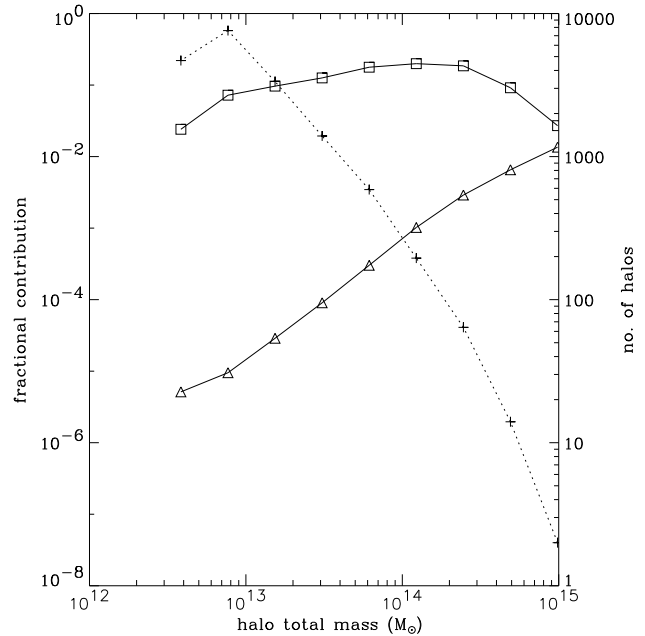,width=0.5\textwidth}}
\caption{As Fig.~\ref{figlambda2cpho}, but for the `W' simulation
  in the $\Lambda$CDM case. The plot shows halo contributions to the
  0.25 keV background when the luminosities of groups are reduced to
  match the extrapolated $L_{\rm X}-T$ relation from White et al.\ (1997).
  By doing so, the background gradually becomes dominated by halos at
  $\sim 10^{14}$\Ms. In the `PW' simulation the luminosities of most
  groups are reduced much further, but the 0.25 keV background does not drop
  much as it becomes `supported' by halos of order $10^{14}$\Ms.}
\label{figlambda7cpho}
\end{figure}

For the last two simulations listed in Table~\ref{tabsims}, we fixed
the luminosities of halos according to observed $L_{\rm X}-T$
relations, as described in section~\ref{secemp}. The resulting
spectra, shown in Figs.~\ref{figopenspec} and \ref{figlambdaspec},
satisfy our upper limit on the 0.25 keV background. In addition, the 1
keV background in the OCDM case is very close to 1 \bgunit, as
expected from observations (see above). (In the $\Lambda$CDM case, the
1 keV background is slightly smaller.)  In spite of the great
difference in the $L_{\rm X}-T$ slopes adopted for groups, the `W' and
`PW' simulations produce spectra that differ by less than a factor of
1/3.  The reason for this is illustrated in Fig.~\ref{figlambda7cpho}
and discussed in the caption. Notice that even in the `W' case,
the contributions of individual groups at 0.25 keV have been reduced
by an order of magnitude or more, across the entire mass range. In a
further simulation, we took the $L_{\rm X}-T$ relation from the `PW'
simulation and `pivoted' the slope below the intersection point at 0.8
keV, to give $L_{\rm X}\propto T^2$ (so that luminosities are higher
than in the `W' simulation). This increased the spectrum around 0.25
keV by a factor of 1/3 relative to the `W' simulation.

\subsection{Source counts}

\begin{figure}
\centerline{\psfig{figure=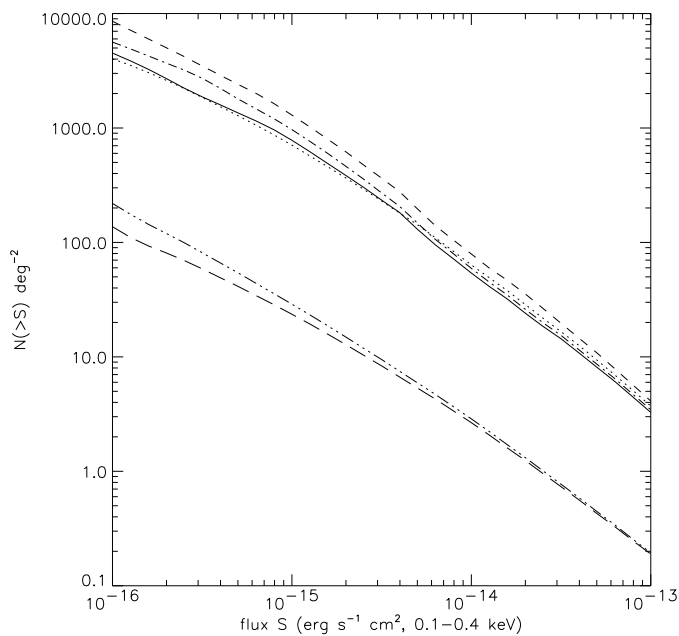,width=0.5\textwidth}}
\caption{Simulated $\log N$--$\log S$ functions in the 0.1--0.4 keV
  band, for the OCDM cosmology. The legend for the different lines is
  given in Table~\ref{tabsims}.}
\label{figopencns}
\end{figure}

\begin{figure}
\centerline{\psfig{figure=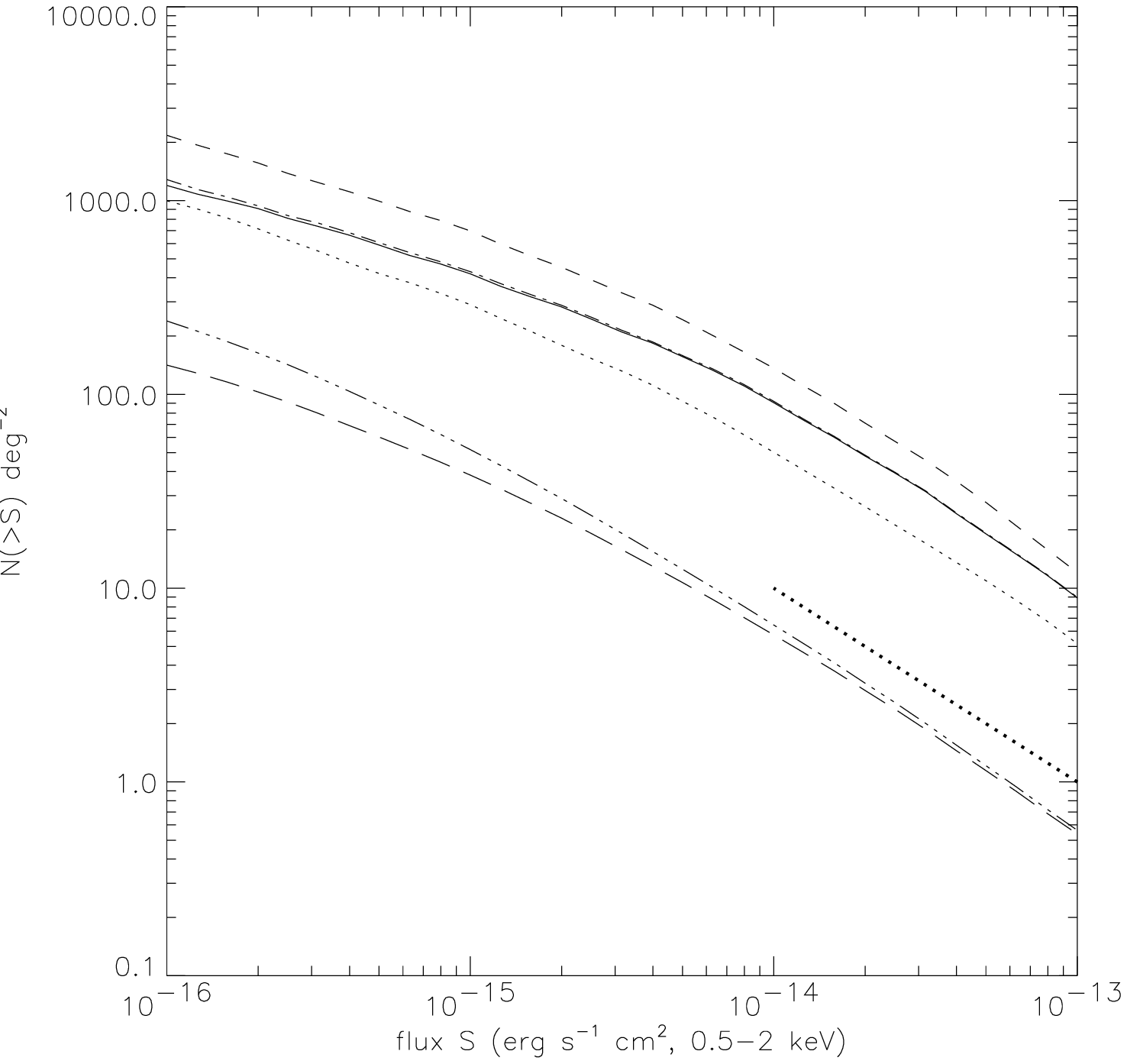,width=0.5\textwidth}}
\caption{As Fig.~\ref{figopencns}, but showing source counts in the
  0.5--2 keV band (OCDM cosmology). The heavy dotted-line approximates
  the observed source counts.}
\label{figopensns}
\end{figure}

\begin{figure}
\centerline{\psfig{figure=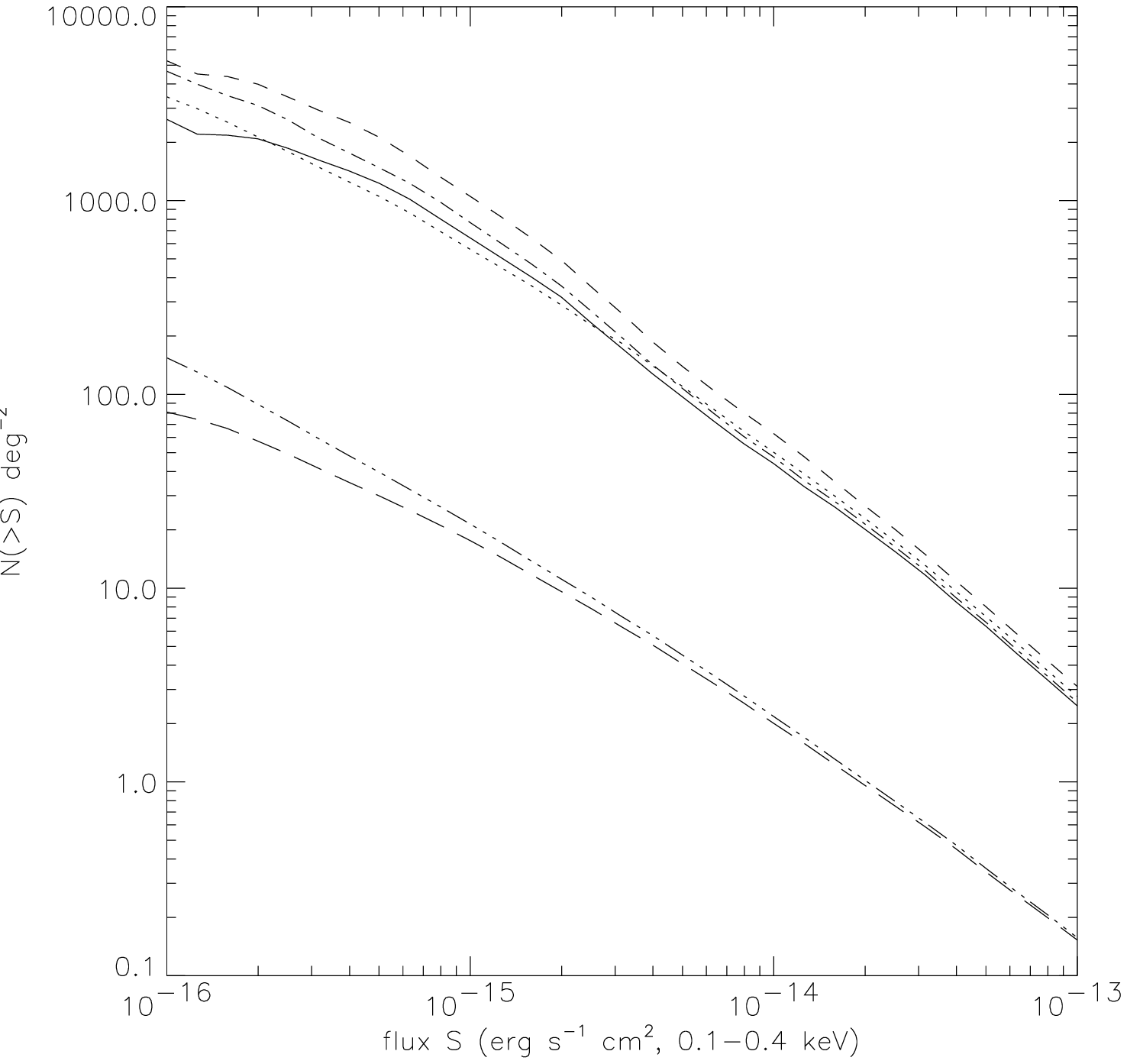,width=0.5\textwidth}}
\caption{Simulated $\log N$--$\log S$ functions in the 0.1--0.4 keV
  band, for the $\Lambda$CDM cosmology. The legend for the different lines is
  given Table~\ref{tabsims}.}
\label{figlambdacns}
\end{figure}

\begin{figure}
\centerline{\psfig{figure=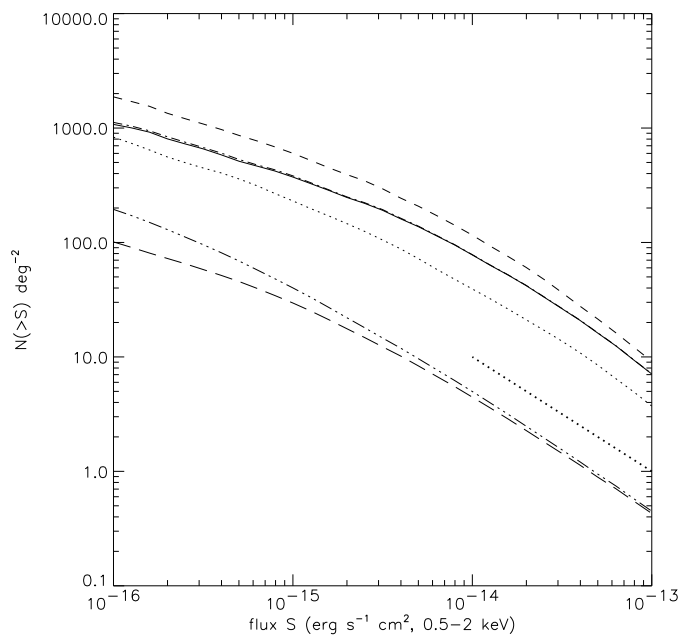,width=0.5\textwidth}}
\caption{As Fig.~\ref{figlambdacns}, but showing source counts in the
  0.5--2 keV band ($\Lambda$CDM cosmology). The heavy dotted-line
  approximates the observed source counts.}
\label{figlambdasns}
\end{figure}

The source counts from the above simulations are displayed in
Figs.~\ref{figopencns} to \ref{figlambdasns}.  Figs.~\ref{figopencns}
and \ref{figopensns} show results from the OCDM simulations in the
0.1--0.4 keV and 0.5-2 keV bands, respectively. Likewise, results for
$\Lambda$CDM are shown in Figs.~\ref{figlambdacns} and
\ref{figlambdasns}.

As with the spectra, the $\log N$--$\log S$ curves fall clearly into
two groups in each figure, given by the first four and the last two
simulations listed in Table~\ref{tabsims}. Little data exists for
comparison, except above $10^{-14}$ \ergpspsqcm\ in the harder band.
The $\log N$--$\log S$ function from the WARPS survey \cite{jsepw98}
extends down to $6\times 10^{-14}$ \ergpspsqcm, and the $\log
N$--$\log S$ function of \citeN{rdbng95} covers fainter fluxes down to
$1\times 10^{-14}$ \ergpspsqcm. Both observed functions are closely
approximated by the simple equation $N(>S)=(10^{-13}
\mbox{\ergpspsqcm}/S)$, which lies an order of magnitude below the
`self-similar' predictions shown in Figs.~\ref{figopensns} and
\ref{figlambdasns}. This is to be expected, for the sources in this
flux range are identified as small clusters in the simulation, with
temperatures of $\sim 1$ keV. Hence, we already know that they should
not be self-similar, as they lie at (or just off) the lower end of
the $L_{\rm X}-T$ relation of \citeN{wjf97}.

Several of the differences between the $\log N$--$\log S$ functions can
be traced back to the spectra. For example, the dotted and solid
curves (which differ in the metallicity used) are very close in the
0.1--0.4 keV band, but differ by almost a factor of 2 in the
0.5--2 keV band. As discussed above, this can be attributed to the iron L
complex.  Reducing $\tau_0$ to 0.1 (dot-dashed curves)
increases the number of low-mass halos with hot gas; this makes
negligible difference in the 0.5--2 keV band, but some change can be
seen in the 0.1--0.4 keV band. In all cases, increasing the CDM shape
parameter to $\Gamma=0.25$ increases the number counts by roughly 50
per cent, though this depends somewhat on the flux.

The $\log N$--$\log S$ functions from the `W' and `PW' simulations
differ little from each other, especially above $10^{-14}$
\ergpspsqcm. This is because the counts are mostly dominated by (small)
clusters in both cases. Therefore they do not depend on which $L_{\rm
  X}-T$ relation we use for groups. This suggests that the predictions
in the 0.5--2 keV band above $10^{-14}$ \ergpspsqcm\ should come close
to the observed counts, since we have imposed the observed $L_{\rm
  X}-T$ relation for clusters and fitted the observed cluster
temperature function. However, the model source counts are a factor of
2 to 3 smaller.  The main reason for this discrepancy is because we do
not include any scatter in the $L_{\rm X}-T$ relation used in the
simulations.  The XRB is not sensitive to scatter, for it measures the
mean properties of halos.  However, scatter in the flux of sources,
when combined with the large number of faint sources compared to
bright ones, can significantly increase the model $\log N$--$\log S$
function.

We now describe a simple correction for this effect. We first note
that the best-fit $L_{\rm X}-T$ relation of \citeN{wjf97} is
calculated in logarithmic space. Therefore the simplest way to include
scatter is to give $\log L_{\rm X}$ (at each temperature) a gaussian
distribution centred on the value given by the best-fit. Let the
standard deviation, $\sigma$, be independent of temperature (see their
Fig.~1a). It follows that the new $\d N/\d S$ distribution is given by
convolving the old one with this gaussian, with $\log S$ as the
independent variable.  We now approximate the model $\log N$--$\log S$
functions (for $S>10^{-14}$ \ergpspsqcm) by the power law
$N(>S)\propto 1/S$---as we did for the observed source counts---which
implies $\d N/\d S\propto 1/S^2$.  It is not hard to show that $\d
N/\d S$ remains a power law when convolved with the gaussian, but it
is shifted upwards by a factor of $10^{2\sigma^2 \ln 10}$. Therefore
an intrinsic scatter in $\log L_{\rm X}$ of $\sigma=0.3$ (\ie a factor
of 2 in each direction) increases $N(>S)$ by a factor of 2.6.
This brings the
model $\log N$--$\log S$ functions roughly in agreement with the data.
However, the true intrinsic scatter for small clusters is not well
determined, and this result highlights the inherent uncertainty in
modelling source counts. This correction should also be applied to the
0.1--0.4 keV band where the slope of $N(>S)$ is the same. At much
lower fluxes, a sizable contribution from groups is not ruled out (see
Figs.~\ref{figopencns} to \ref{figlambdasns}). If their scatter is
very large (as discussed below), then their counts would be increased
in the same way.

\section{Further investigations} \label{secfur}

\subsection{Non-gravitational heating while maintaining the same gas
  fractions} \label{secmarginal}

Here we discuss an example of heating that fails remarkably to reduce
the simulated XRB to the required level.

In our heating models described in WFN99, we allow the total energy of
gas halos (the sum of thermal and potential energies) to increase in
the presence of excess energy from non-gravitational heating. This
both raises the gas temperature and flattens the density profile of a
gas halo, but we maintain the same gas mass within $r_{\rm vir}$. For
the isothermal gas profiles used in this paper, this is modelled by
reducing the slope parameter $\eta$. The X-ray luminosity can
be reduced by up to an order of magnitude as a result, before the
total energy of the gas halo becomes positive (relative to the
appropriate potential), at which point we regard the gas as
gravitationally unbound.

\begin{figure}
\centerline{\psfig{figure=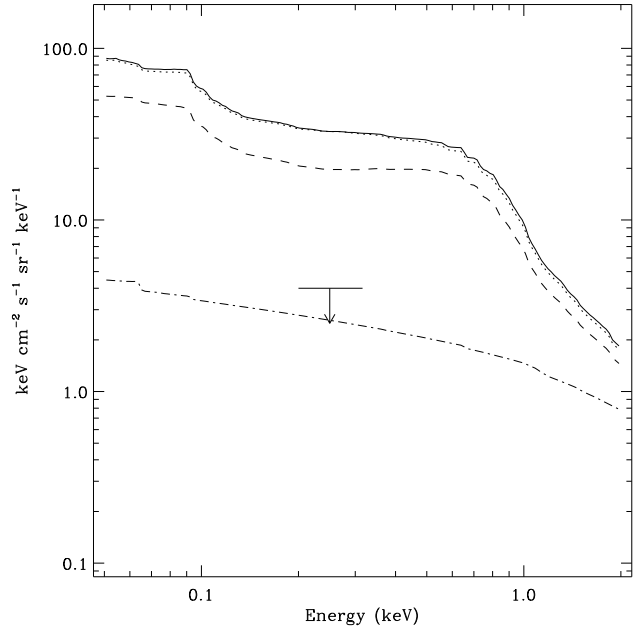,width=0.5\textwidth}}
\caption{Further simulated spectra in the OCDM cosmology.
  The solid spectrum from Fig.~\ref{figopenspec} is shown here for
  comparison. The dashed spectrum was obtained by heating all gas
  halos to the point of being marginally bound (see
  section~\ref{secmarginal}). For the dotted spectrum, the gas
  fractions of halos were determined naturally from their progenitors
  (section~\ref{secinherit}). The dot-dashed spectrum was given by a
  preliminary study of the injection of excess energy by galaxies,
  which simultaneously fits the $L_{\rm X}-T$ relations for clusters
  (section~\ref{secprelim}).}
\label{figopenother}
\end{figure}

\begin{figure}
\centerline{\psfig{figure=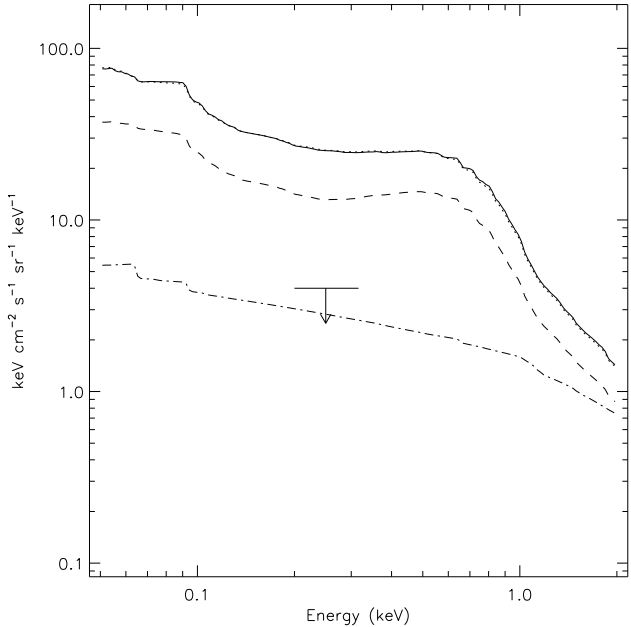,width=0.5\textwidth}}
\caption{As Fig.~\ref{figopenother}, but for $\Lambda$CDM.
  (The solid spectrum is taken from Fig.~\ref{figlambdaspec}.)}
\label{figlambdaother}
\end{figure}

In two further simulations, we modify the first model in
Table~\ref{tabsims} by heating the gas in all halos $>3\times
10^{12}$\Ms\ to the point of being marginally bound, whilst retaining
a gas fraction of 0.17.  (100 realisations were used in these
simulations.)  The strong heating roughly doubles the temperature of a
halo and halves the value of $\eta$.  These simulations may be
expected to give the maximum possible reduction in the XRB given a gas
fraction of 0.17. However, the resulting XRB spectra, given by the
dashed curves in Figs.~\ref{figopenother} and \ref{figlambdaother},
show a reduction of only a half (at 0.25 keV) compared to the solid
spectrum.

\begin{figure}
\centerline{\psfig{figure=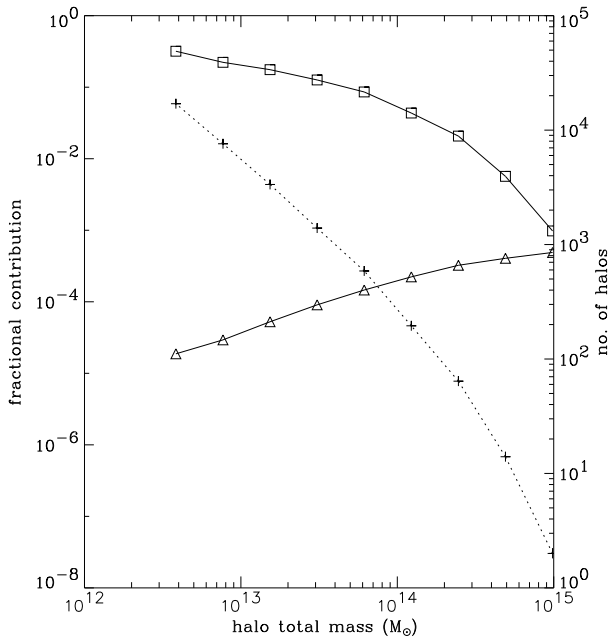,width=0.5\textwidth}}
\caption{Plot of halo contributions to the 0.25 keV background 
  when gas halos are maximally heated as
  described in section~\ref{secmarginal}, for the $\Lambda$CDM case.
  The symbols are described in Fig.~\ref{figlambda2cpho}.}
\label{figlambda8acpho}
\end{figure}

Notice that we have not heated halos below $3\times 10^{12}$\Ms. Doing
this doubles the dashed spectrum at 0.25 keV, due to a whole new
contribution from halos of $\sim 10^{12}$\Ms---based on the criterion
of $\tau_0=1$, these halos become hot collapses (see
section~\ref{sectau}) as a result of their increased cooling times. However,
we are more interested in how much we can {\em reduce} the XRB.

Fig.~\ref{figlambda8acpho} shows the new distribution of halo
contributions for $\Lambda$CDM. Except for the point at $4\times
10^{12}$\Ms, the total contributions (squares) at all higher masses
are reduced by the heating. (The contribution at $4\times 10^{12}$\Ms\ 
increases for the same reasons given above, but the effect on the
total 0.25 keV background is small.)  The reason for the rather modest
reduction in the 0.25 keV background has to do with the large fraction
of gas able to cool in the smaller groups when heating is absent. The
fraction of gas that is able to cool decreases gradually as we
progress to massive groups. The depletion of gas from the centres of
halos can greatly reduce the luminosity, so that even when heating is
absent the groups do {\em not} follow $L_{\rm X}\propto T^2$ on
average, but obey a steeper relation. When the halos are heated in the
manner described, the fraction of gas able to cool becomes small even
in small groups, and we recover $L_{\rm X}\propto T^2$ (because
$\eta\approx 5$ in all groups) normalized to a lower luminosity.
Thus, the modest reduction of the XRB may be attributed to the
already-reduced luminosities of groups (averaged over time) due to
cooling.  
This also explains why heating reduces the contributions of large
groups more than for small groups, as shown by the sloping curve in
Fig.~\ref{figlambda8acpho} (cf.\ Fig.~\ref{figlambda2cpho}). 

Note that if we impose $L_{\rm X}\propto T^2$ for all groups and
clusters when heating is absent, then this corresponds to ignoring the
effect of cooling.  As we mentioned in the previous section, this
results in a 0.25 keV background that is 5 times higher than our
fiducial result, and therefore about 10 times higher than the dashed
spectra.

The inability of this method to reduce the 0.25 keV background to a
level that is even close to the upper limit, implies that the gas
fractions of groups (before the gas cools) must be much lower than
0.17. In principle, we can remove, say, all of the gas in groups below
$2\times 10^{13}$\Ms\ (see Fig.~\ref{figlambda8acpho}), which would bring
the 0.25 keV background just below the upper limit; but the model is
already so contrived that groups are almost certainly gas-poor up to
$\sim 10^{14}$\Ms. This conclusion is supported by
results from X-ray studies of individual groups.  Since the average
gas fraction of clusters is at least 0.17, we argue that
a large fraction of the gas belonging to groups must be outside their
virial radii (see below).

\subsection{Gas fractions determined by inheritance} \label{secinherit}

So far, we have fixed the initial gas fractions of groups at 0.17.  We
now consider the effect of relaxing this assumption, so that the
amount of gas in a halo is naturally determined by the amount accreted
or inherited from its progenitors. We use the same parameters as the
first model in Table~\ref{tabsims}, but modify the primordial gas
fraction so that we obtain large clusters with gas fractions close to
0.17. In the OCDM case, we use a primordial gas fraction of 0.25, and
for $\Lambda$CDM we use 0.23. ($\Gamma$ is increased marginally by the
new values of $\Omega_b$, according to \citeNP{sugiy95}.)  As in the
fiducial simulations, we assume no non-gravitational heating. 100
realisations were used in each simulation.  The resulting spectra are
given as dotted curves in Figs.~\ref{figopenother} and
\ref{figlambdaother}.

The dotted spectra are surprisingly close to the fiducial results and
almost coincide with them. This in fact hides a large scatter in the
gas fractions of individual groups. For example, in halos of $8\times
10^{12}$\Ms\ we find that initial gas fractions of 0.14 to 0.20 are
common. Star formation was included in galaxies according the model
described in WFN99.
This consumes gas at the 10 per cent level (using
clusters as samples of baryons), but hot gas that cools in groups is
converted into baryonic dark matter (see section~\ref{sectau}).
The effect of excess energy from supernova heating on the
gas halos of groups and clusters is ignored in these simulations.

Thus as far as the XRB is concerned the primary difference from our
fiducial simulations is the freeing of the gas fractions. Although a
large amount of gas is able to cool in a small group, this is
compensated at the next collapse by new material with relatively high
gas fractions. As a result, the average gas fraction of
newly-collapsed groups remains close to 0.17.
Intuitively, one might expect the groups to have higher initial gas
fractions than the clusters, as would be the case if clusters formed
only from group-group mergers, but this is of course not true in
reality. These results show that the simplification made in our
fiducial simulations has remarkably little effect on the predicted XRB.

An improvement on our model would be to follow the growth of halos
more closely in time, by refining the simulation of merger trees.
However, it seems that adding gas more continuously to halos (compared
to adding it in one go when a larger block is ready to collapse) would
be more likely to increase the predicted XRB than to reduce it.

\subsection{Energy injection from galaxies} \label{secprelim}
In section~\ref{secmarginal} we found that the gas fraction in
collapsing groups must be much lower than that in clusters, but from
the discussion in section~\ref{secinherit} it is clear that cooling
alone cannot produce this result.
There must be other mechanisms to prevent gas from following the dark
matter into the halos of groups, but which do not prevent it from
falling into rich clusters.

Such a scenario can be achieved by giving the gas sufficient excess
energy so that it would not be gravitationally bound to a group, but
would succumb to the deeper potential well of a cluster.

The minimum excess energy required to unbind a halo is that required
to make its total energy zero relative to the appropriate potential.
A scatter plot of this minimum energy as a function of halo mass is
given in Fig.~7 of WFN99 (where it is called the `binding energy' of
the halo). However, the current work only demonstrates that it is
necessary to drive the bulk of the gas beyond $r_{\rm vir}$ and this
may require significantly less energy (see also \citeNP{bbp99}).  Some
clouds may be heated more strongly than others; while the bulk of the
gas may be unbound, some may still collapse to form a halo with a much
lower gas fraction.  This complicates any estimate of the minimum
excess energy required to explain the 0.25 keV background.  However,
if the bulk of non-gravitational heating is injected in the smaller
`branches' of a merger tree ($\la 5\times 10^{12}$\Ms) by the galaxies
within, then the excess energy of gas associated with groups should be
around 1--3 keV/particle (or more to allow for dilution), since this
is the excess energy required to fit the properties of X-ray clusters
(WFN99).  This is similar to a model proposed by Pen (1999).

A large energy input is unlikely to be uniform, so that one
consequence of this scenario would be a large increase in the scatter
of the $L_{\rm X}-T$ or the $L_{\rm X}-\sigma$ distribution as we go
below $\sim 1$ keV and gas halos become (partially) unbound. An
increase in the scatter of these distributions is suggested by the
Hickson Compact Groups \cite{pbeb96}. Indeed the large scatter in
properties also extends to the $T-\sigma$ distribution (see also
Helsdon \& Ponman 1999).

In the final simulations we make a preliminary study of the injection
of excess energy in halos below $10^{12}$\Ms. We assume that the
energy is due to AGN and possibly supernovae. The energy is then
retained in the gas as long as it is not radiated. In our model no
excess energy is lost while the gas remains outside virialized halos (see
also below). In order to simultaneously fit the constraint on the XRB
and the $L_{\rm X}-T$ relation for clusters, we require an excess
specific energy of around 2.8 keV/particle in cluster gas (WFN99,
Model B for isothermal profiles only). This flattens the density
profiles of small clusters and reduces their luminosity. As discussed
in our earlier paper, the excess energy in the cluster only
approximates the actual energy injection required, which is likely to
be smaller because a `gravitational contribution' to the excess energy
can result when the gas is displaced by strong heating.  Nevertheless,
this is the approximation made in the simulations.

Our prescription for energy injection is as follows: we give all gas
that was ever associated with a halo in the range
(0.015--1)$\times 10^{12}$\Ms\ an excess specific energy of 7
keV/particle (in the OCDM case) or 5 keV/particle ($\Lambda$CDM).  In
practice, we simply set the excess energy of gas to this level at the
end of the relevant collapse steps; this
occurs even if the gas is not bound to the halo---it only needs to be
associated with the dark matter in the halo.  Due to dilution by
unheated gas, these result in excess energies of around 3
keV/particle in clusters, but with a significant scatter.
We also lower the
primordial gas fraction to 0.18, but all other parameters are the same
as in section~\ref{secinherit}.  Since we do not model partially
unbound halos, most groups have no gas at all in the simulations.  We
used 400 realisations for each run.

The resulting spectra are shown as dot-dashed curves in
Figs.~\ref{figopenother} and \ref{figlambdaother}. They satisfy the
upper limit in both cases.  More dilution of
the excess energy seems to occur in the OCDM cosmology, which
therefore required a higher `initial' excess specific energy.

The resulting clusters have gas fractions of about 0.17. By comparing
with the primordial gas fraction it is evident how little gas has
cooled. Although we expect hardly any gas to cool in groups, galaxies
are also strongly affected by the excess energy from their
progenitors. In the simulations, this is exacerbated by the averaging
of excess energies over all the gas associated with a collapsing halo.
To avoid this problem, it may be necessary to inject most of the
required excess energy in halos comparable to $\sim 10^{12}$\Ms.  

As a result of heating, gas would be ejected from a halo as a wind,
terminating any star formation in the process. Such a wind may
naturally result from the growth of a massive black hole (Fabian
1999)\nocite{fabia99}. To estimate the black hole growth that is
required, let the energy available for the heating of gas be given by
$\epsilon\: \Delta M_{\rm BH} c^2$, where $\Delta M_{\rm BH}$ is the
mass accreted onto the black hole and $\epsilon$ is the mass-to-energy
conversion rate. If we let $\epsilon=0.1$ (the value often used for
the mass-to-light conversion rate), and distribute the energy over gas
of mass $M_{\rm gas}$, then we obtain an excess specific energy equal
to $0.1 c^2 \Delta M_{\rm BH}/M_{\rm gas} = 5.8\times 10^4 (\Delta
M_{\rm BH}/M_{\rm gas})$ keV/particle. Thus, $10^7$\Ms\ of black hole
growth is able to supply $10^{11}$\Ms\ of gas with about 6
keV/particle. In a similar calculation in WFN99, we used the black
hole density of the universe to show that about 4 keV/particle of
excess energy can be obtained in this way even if it is averaged over
all the baryons in the universe.

Once the gas is heated above a temperature of 1 keV in a galaxy halo,
its cooling time becomes very long---comparable to a Hubble time,
depending on its density. As the gas expands, its cooling time remains
high ($t_{\rm cool}\propto 1/(T^{1/2}\Lambda(T,Z))$ if the gas expands
adiabatically), but more importantly it converts a large fraction of
its thermal energy into potential and kinetic energies. In this way,
the gas should be able to retain most of its excess energy until it
recollapses into a cluster.

\section{Implications for the IGM and galaxy formation} \label{secimp}

In the previous section, we first showed that the group population as
a whole has a much lower gas fraction than X-ray clusters, and then
proposed non-gravitational heating at the level of $\sim 1$
keV/particle as a means of accounting for both
the properties of groups and X-ray clusters, in a natural and
self-consistent manner. Indeed, the low gas fractions of groups and
the large intrinsic scatter in their properties compared to clusters
(section~\ref{secprelim}) both lend support to this high level of heating,
which we originally proposed for clusters.  
Similarly, \citeN{renzi97} noted a precipitous drop in the
intra-cluster medium (ICM) mass-to-light ratio at around 1 keV, as
well as in the ICM iron mass-to-light ratio at the same temperature
(`light' refers to the total B-band luminosity of the galaxies in the
cluster).  He concluded that clusters could not have formed by
assembling groups similar to those observed. In the heating scenario,
this is resolved by allowing the gas to `reunite' with the groups on
the formation of a cluster.

In the proposed model, the high energy of heated gas in the IGM means
that it cannot remain in the filamentary and sheet-like structures
seen in N-body simulations. The actual distribution of the different
gas phases in the IGM would depend on details of how heating occurred,
a problem analogous to supernova heating in the inter-{\em stellar}
medium.  This is contrary to the evolution of baryons described by
\citeN{co99}.  In their cosmological hydrodynamic simulations, almost
half of the baryons at redshift zero lie in the temperature range
0.01--1 keV and exist in filaments and more clumpy structures. This
gas is heated primarily by collisions, where the energy has a
gravitational origin. When heating is included at the proposed level,
a large fraction of this gas would be expelled from the potential
wells of filaments and groups, resulting in a smoother and more
diffuse gas distribution.  As a result, the likelihood of detecting
this large reservoir of baryons in the universe would be much reduced
\cite{fhp98}.

Since even the smallest groups are affected by the heating, it seems
likely that the heating process would play an important part in the
evolution of galaxies.  For example, the high end of the luminosity
function of galaxies (where the Schechter function decays
exponentially) may be influenced by the heating process. Assuming that
the source of energy is AGN, the possibility that AGN and galaxy
formation are intimately connected will continue to be a topic of much
discussion.

\section{Summary of conclusions} \label{secconc}

We have made a systematic study of the soft X-ray background from hot
gas halos, with the aim of constraining the properties of the gas
halos of groups. Using Monte Carlo simulations of halo merger trees
coupled with realistic gas density profiles, and including the effects
of gas removal due to cooling, we calculated the XRB spectrum along with
source counts in the 0.1--0.4 keV and 0.5--2 keV bands. In
addition, we investigated the composition of the XRB in terms of the
masses of groups that contribute at 0.25 keV and 1 keV.
Our main conclusions are as follows:
\begin{itemize}
\item Radio-quiet quasars are able to account for almost all of the
  extragalactic XRB at 0.25 keV. As a result, we set an upper limit of
  4 \bgunit\ on the 0.25 keV background due to gas halos.
\item In the absence of non-gravitational heating, the predicted 0.25 keV
  background is an order of magnitude higher than this upper limit.
  In addition, it is contributed by the entire mass range of groups, 
  from $\sim 5\times 10^{12}$ to $\sim 10^{14}$\Ms.
\item The removal of gas due to cooling plays an important part in
  determining the XRB in this case. Excluding this effect increases
  the predicted 0.25 keV background by about a factor of 5.
\item Maximally heating the gas halos of groups without changing their gas
  fraction reduces the 0.25 keV background by only a factor of two or
  less. It follows that most of the gas associated with groups, down
  to halos of $\sim 5\times 10^{12}$\Ms, must be outside the virial
  radii of these halos.  
\item The properties of both groups and X-ray clusters can be
  naturally explained by a model in which the gas is given (on
  average) excess specific energies of 1--3 keV/particle by
  non-gravitational heating.
\item In addition to satisfying the constraint on the XRB, this would
  result in a large scatter in the X-ray properties of groups
  (assuming that the heating is inhomogeneous), as well as a dichotomy
  in the properties of gas halos above or below $T\sim 1$ keV, both of
  which are supported by observations.
\item The excess energy is most likely injected by galaxies in
  the smaller `branches' of the halo merger tree ($\la 5\times
  10^{12}$\Ms), by active galactic nuclei and possibly supernovae.
\item This greatly reduces the likelihood of detecting the large
  reservoir of cosmic baryons which would otherwise be expected in
  groups and filaments.

\end{itemize}

\section*{Acknowledgements}
KKSW thanks Omar Almaini for helpful advice, and is
grateful to the Croucher Foundation for support. ACF thanks the Royal
Society for support.
PEJN gratefully acknowledges the hospitality of the
Harvard-Smithsonian Center for Astrophysics. This
work was funded in part by NASA grants NAG8-1881, NAG5-3064 and
NAG5-2588.

\bibliography{mnrasmnemonic,paper,bookastrophys,bookcosmology}
\bibliographystyle{mnras}

\end{document}